\documentclass[aip,jap,reprint]{revtex4-1}
\usepackage{graphicx}
\pdfoutput=1

\begin{document}

\widetext
\leftline{The following article has been submitted to/accepted by Applied Physics Letters.}
\leftline{After it is published, it will be found at \url{http://apl.aip.org/}.}

\title{Spin lifetime measurements in GaAsBi thin films}%

\author{Brennan Pursley}%
\email{bpursley@umich.edu}
\affiliation{Applied Physics Program, University if Michigan, Ann Arbor, MI 48109}

\author{M. Luengo-Kovac}
%\email{}%
\affiliation{Department of Physics, University of Michigan, Ann Arbor, MI 48109}

\author{G. Vardar}
%\email{}%
\affiliation{Department of Materials Science and Engineering, University of Michigan, Ann Arbor, MI 48109}

\author{R. S. Goldman}
%\email{}%
\affiliation{Applied Physics Program, University if Michigan, Ann Arbor, MI 48109}
\affiliation{Department of Physics, University of Michigan, Ann Arbor, MI 48109}
\affiliation{Department of Materials Science and Engineering, University of Michigan, Ann Arbor, MI 48109}

\author{V. Sih}
%\email{}%
\affiliation{Applied Physics Program, University if Michigan, Ann Arbor, MI 48109}
\affiliation{Department of Physics, University of Michigan, Ann Arbor, MI 48109}

\date{November 8, 2012}%
%\revised{August 2010}%

\begin{abstract}
Photoluminescence spectroscopy and Hanle effect measurements are used to investigate carrier spin dephasing and recombination times in the semiconductor alloy GaAsBi as a function of temperature and excitation energy.  Hanle effect measurements reveal the product of g-factor and effective spin dephasing time ($gT_s$) ranges from 0.8 ns at 40 K to 0.1 ns at 120 K.  The temperature dependence of $gT_s$ provides evidence for a thermally activated effect, which is attributed to hole localization at single Bi or Bi cluster sites below 40 K.
\end{abstract}

%\keywords{none}%

\maketitle

The field of spintronics, firmly entrenched in giant magnetoresistive (GMR) read heads for hard-drive technology, is rapidly progressing to encompass other elements of computer memory such as magnetic random access memory (MRAM).\cite{Keatley2011a}  However, a superior spintronic analogue to silicon compimentary metal-oxide semiconductor (CMOS) logic systems has yet to be realized.  The dilute bismuthides GaAs$_{(1-x)}$Bi$_x$, also referred to as bismides, are a potentially promising family of semiconductor alloys for both spintronic and electronic applications.   These alloys can be grown on common GaAs substrates, Bi incorporation has minimal effect on electron mobility,\cite{Kini2009,Cooke2006a} and by incorporating nitrogen, the significant band gap tunability of both bismuthide and nitride alloys might be achieved while remaining lattice matched to GaAs.\cite{Tiedje2008}  Furthermore,  giant spin-orbit bowing in films with low Bi incorporation has been reported\cite{Fluegel2006} which allows tuning of the spin-orbit splitting.  Although bismuthide alloys have intriguing properties that would be useful for scalable spintronic devices, the carrier spin dynamics in GaAsBi have not yet been reported.

In this work, we investigate spin dephasing and carrier recombination times in dilute bismuthide thin films using photoluminescence spectroscopy and Hanle effect measurements.  Photoluminescence (PL) below 1.4 eV is attributed to carrier recombination in the GaAsBi epilayer and we observe a power dependent blue shift of the emission peak in agreement with existing literature.\cite{Kudrawiec2009a,Imhof2010,Francoeur2008}   In addition, we report excitation dependent broadening of the spectra between the main bismuthide and GaAs emission.  Hanle effect measurements reveal the product of g-factor and effective spin dephasing time ($gT_s$) ranges from 0.8 ns at 40 K to 0.1 ns at 120 K with $gT_s$ nearly constant below and decreasing above 40 K.  All observed phenomena are attributed to hole localization that occurs below 40 K at single Bi or Bi cluster sites.

GaAsBi epilayers were grown by molecular-beam epitaxy on GaAs substrates at 350º C  with a growth rate of 0.1 $\mu$m/hr and As$_2$/Ga and Bi/As$_2$ beam equivalent pressure ratios of 6 and 0.01, respectively.  A 100 nm thick GaAs$_{0.992}$Bi$_{0.008}$ film was grown on a 520 nm GaAs buffer layer on a semi-insulating (001) GaAs substrate.  The GaAsBi film composition was confirmed using Rutherford back-scattering spectroscopy.

Samples were mounted on the cold-finger of a continuous flow liquid helium cryostat.  PL was performed using a tunable-wavelength mode-locked Ti:Sapphire laser with pulse duration $\sim$3 ps and a repetition rate of 76 MHz as the excitation source.  Two excitation energies, ExA=1.59eV and ExB=1.45eV, were used to selectively excite carriers either above (ExA) or below (ExB) the observed GaAs band gap at all temperatures.  The excitation and collection paths were both normal to the sample surface.  Incident light was focused to a 75 $\mu$m diameter spot with intensities ranging from 1 W/cm$^2$ to 250 W/cm$^2$.  A grating spectrometer and liquid nitrogen cooled CCD camera with 0.2 nm resolution were used for analysis.  Using both excitation energies over the entire intensity range, PL spectra were recorded at temperatures between 10 K and 200 K.  Laser scatter was reduced by orthogonally oriented linear polarizers and long pass filters.

We measured the Hanle effect using either ExA or ExB excitation, with an irradiance of 23 W/cm$^2$.  A variable retarder, placed in the incident path, allowed for right ($\sigma_+$) or left ($\sigma_-$) circularly polarized excitation.  A quarter-wave plate, combined with a linear polarizer and placed in the collection path, selectively passed $\sigma_+$ PL.  We recorded PL within the range from 1.30 eV to 1.38 eV, and in our analysis, we present Hanle data averaged within a $\pm2$ meV range of selected energies.  An electromagnet was used to generate a field with maximum magnitude of 250 mT in the sample plane.  For each measurement, the applied magnetic field was varied from -250 mT to +250 mT.  At each field step, the $\sigma_+$ PL intensity was recorded for both $\sigma_+$ and $\sigma_-$  excitation.  The entire process was repeated several times to average out noise due to fluctuations in the laser power.  The relative polarization $P_{rel}=(I_{\sigma_+}-I_{\sigma_-})/(I_{\sigma_+}+I_{\sigma_-})$, where $I_{\sigma_j}$ is the intensity of $\sigma_+$ PL under $\sigma_j$ excitation, was calculated using the average intensity of a $\pm$2 meV range about selected PL energies.

The polarization of PL in direct-gap III-V semiconductors is proportional to the spin polarization of the photoexcited carriers.  Equation \ref{gov} describes the temporal evolution of electron spin polarization where $\Omega$ is the Larmor precession frequency, $\tau_D$ is the ensemble dephasing time, $\tau_R$ is the carrier recombination time, and $S_0$ is the initial spin polarization.\cite{Meier1984}
\begin{equation}
\label{gov}
\frac{d\mathbf{S}}{dt}= \mathbf{\Omega} \times \mathbf{S} -\frac{\mathbf{S}}{\tau_D} -\frac{\mathbf{S}-\mathbf{S}_0}{\tau_R}
\end{equation}
In bulk GaAs, the degeneracy of the light and heavy hole transitions lead to $S_0$=0.5 for transitions involving circularly-polarized light near k = 0.\cite{Zutic2004a}  However, $S_0$ could have a different value in bismuthide samples due to carrier trapping\cite{Francoeur2008} and strain-induced splitting of the valence bands.\cite{Francoeur2003}  Therefore, our only assumption is that $S_0$ is a constant.  $\mathbf{\Omega}$, a function of the applied field, is defined by Eq. \ref{omega} where $g$ is the effective electron g-factor, $\mu_B$ is the Bohr magneton, and $\hbar$ is the reduced Planck’s constant.
\begin{equation}
\label{omega}
\mathbf{\Omega} =\frac{g\mu_B \mathbf{B}}{\hbar}	
\end{equation}
It is convenient to define an effective spin lifetime, $T_S$, by Eq. \ref{Ts}.
\begin{equation}
\label{Ts}
\frac{1}{T_S} = \frac{1}{\tau_D} + \frac{1}{\tau_R}
\end{equation}
In steady state, the combination of Eqs. \ref{gov} through \ref{Ts} yields \ref{steady}.
\begin{equation}
\label{steady}
0=  \frac{g \mu_B}{\hbar} \mathbf{B}\times\mathbf{S} - \frac{\mathbf{S}}{T_S} + \frac{\mathbf{S_0}}{\tau_R}
\end{equation}
The spin polarization, $S_0$, of the photo-generated carriers is defined to be along $+\mathbf{\hat{z}}$, the same direction as the incident light.  By applying a magnetic field $\mathbf{B}$ in the sample plane, which is perpendicular to $\mathbf{\hat{z}}$, we obtain
\begin{eqnarray}
S_z(B) &=&  \frac{S_z (0)}{1+(\frac{gT_s \mu_B}{\hbar}B)^2} \label{SzB} \\
S_z(0) &=& \frac{S_0}{1+\frac{\tau_R}{\tau_D}} \label{Sz0}
\end{eqnarray}
where $S_z(0)$ is the zero-field maximum and $S_z(B)$ has a Lorentzian lineshape.  Note that $gT_s$, the product of the effective spin lifetime and g-factor, cannot be uncoupled through Hanle measurements.  The corresponding equations for the emitted PL polarization are
\begin{eqnarray}
P(B) &=&  \frac{P(0)}{1+(\frac{gT_s \mu_B}{\hbar}(B-B_0))^2}+P_{bkg}    \label{PB} \\
P(0) &=&  \frac{S_0^2}{1+\frac{\tau_R}{\tau_D}} \label{P0}
\end{eqnarray}
The $S_0^2$ in Eq. \ref{P0} accounts for both the optical generation and recombination pathways of spin polarized carriers.  $B_0$ and $P_{bkg}$ are, respectively, horizontal and vertical offsets added for better fitting.

\begin{figure}[t]
\includegraphics[width=8.5cm]{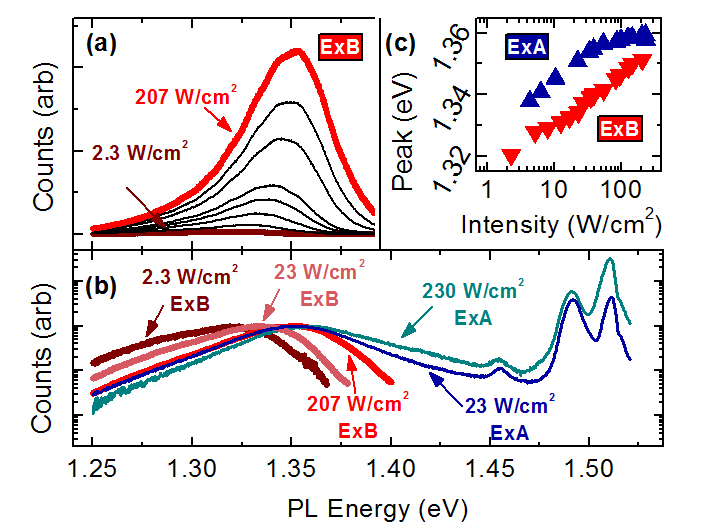}
\caption{a) Power dependent photoluminescence (PL) generated by a Ti:Sapphire laser tuned to 1.45 eV (ExB) for a sample temperature of 10 K.  b) Comparison of PL, normalized to the dilute bismuthide emission peak, for excitations of 1.45 eV and 1.59 eV (ExA) at 10 K and various powers, as labeled.  Red lines are for ExB and blue lines for ExA excitations, respectively.  c) Power dependence of the bismuthide emission peak location at 10 K for both ExB (red down triangles) and ExA (blue up triangles).}
\label{f1}
\end{figure}

Figure \ref{f1}(a) shows PL attributed to carrier recombination in the GaAsBi epilayer measured at 10 K with ExB at intensities of 2.3 W/cm$^2$ to 207 W/cm$^2$.  Increasing power blue shifts the bismuthide emission peak from 1.32 eV to 1.36 eV and increases the low energy tail.  PL measured at 10 K, 40 K, 80 K, and 120 K all exhibit similar power dependence even though PL intensity diminishes with increasing temperature.  Figure \ref{f1}(b) shows a comparison of ExA and ExB generated PL collected at 10 K and normalized to the bismuthide signal maximum, which is below 1.40 eV.  The three peaks observed above 1.40 eV are attributed to emission from the underlying GaAs:  1.46 eV, neutral acceptor exciton; 1.48 eV, valence to conduction band transition; 1.51 eV, free and neutral donor excitons.\cite{Kang1996,Zemon1986}  ExA at 23 W/cm$^2$ yielded roughly the same bismuthide emission peak location as ExB at 207 W/cm$^2$, but with ExA there is a broadening of the bismuthide emission at higher energies.  At all temperatures, excitation with ExA compared to ExB generated a broader high energy tail to the bismuthide signal.  Furthermore, low powers of ExA generated PL with the same bismuthide peak locations as high powers of ExB.  The bismuthide peak locations obtained at 10 K for both excitations as a function of power is shown in Figure \ref{f1}(c).

We attribute the differences in PL linewidth and bismuthide emission peak location between excitation energies to carrier transfer from the underlying GaAs to the dilute bismuthide layer as opposed to sample heating.  Migrating carriers generated in GaAs by ExA would increase the carrier density of the bismuthide layer and fill higher energy states, continuing the power dependent trends for ExB, as observed in \ref{f1}(b-c).  PL measurements show that the bismuthide peak location exhibits minimal shift when the sample temperature is varied from 10 K to 120 K.  Carrier transfer is supported by the theoretical prediction,\cite{Broderick2011,Usman2011} and likely observation,\cite{Riordan2012} of a type I heterojunction at the GaAsBi/GaAs interface which energetically favors both electron and hole migration from GaAs to the bismuthide layer.  The power dependence of the bismuthide signal shift is likely due to the filling of first the defect, single-Bi, and Bi-cluster bound exciton states, followed by the band edges.  Several recent studies support the interpretation of hole localization at Bi and Bi-cluster sites through Bi atom distributions,\cite{Sales2011,Ciatto2008} bandgap transition behavior,\cite{Kudrawiec2009a,Imhof2010,Francoeur2008} hole mobility,\cite{Kini2011} and hole diffusion.\cite{Sales2011}   The measured spin behavior, discussed below, provides further support for the effects of carrier transfer and localization.

\begin{figure}[t]
\includegraphics[width=8.5cm]{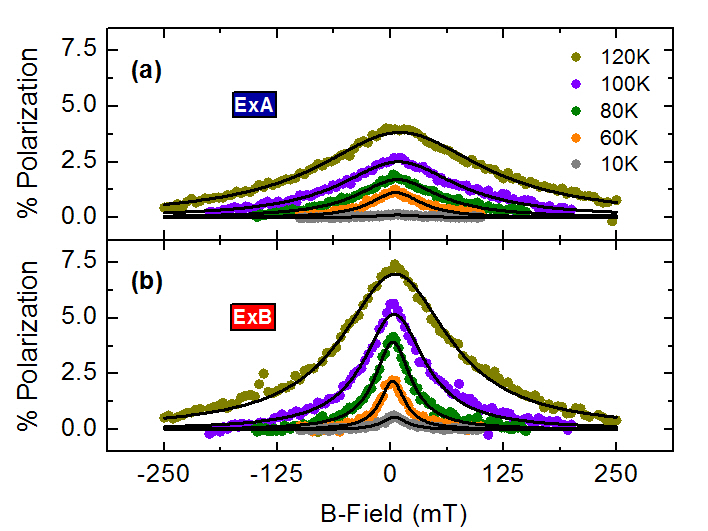}
\caption{Temperature dependent Hanle data for excitation energies of a) 1.59 eV (ExA) and b) 1.45 eV (ExB) evaluated at their respective 10 K bismuthide photoluminescence peak locations of 1.35 eV and 1.33 eV.  Polarization and linewidth monotonically decrease as temperature decreases.}
\label{f2}
\end{figure}

Figure \ref{f2} shows the temperature dependent evolution of Hanle data for PL energies of 1.35 eV with ExA (Fig. \ref{f2}a) and 1.33 eV with ExB (Fig. \ref{f2}b), the respective PL peak energies for each excitation at 10 K.  The data were shifted to remove the magnetic field independent vertical offset $P_{bkg}$ attributed to laser scatter.  Nonlinear least-squares fits were performed to all data with Eq. \ref{PB} and are shown as black lines in Fig. \ref{f2}.  The fitting parameters were $P(0)$, $gT_S$, $P_{bkg}$, and $B_0$.  For both excitation energies, the amplitude and linewidth of the Hanle curves increases with increasing temperatures.  Equation \ref{PB} shows that the value of $gT_S$ has a direct impact on the observed Hanle linewidth--large $gT_S$ leads to small widths and vice versa.  From Eq. \ref{P0}, and assuming that $S_0$ is constant, an increase in polarization corresponds to a decrease in the ratio $\tau_R/\tau_D$.  Since $P(0)$ can obtain at least 7$\%$ at 120 K, this implies that at low temperature, where the polarization has decreased by an order of magnitude, $\tau_R$ is at least an order of magnitude larger than $\tau_D$.  According to Eq. \ref{Ts}, a very large $\tau_R/\tau_D$   value implies $gT_S\simeq g\tau_D$.  Therefore $g\tau_D$ dominates the observed low temperature Hanle linewidth.
 
\begin{figure}[t]
\includegraphics[width=8.5cm]{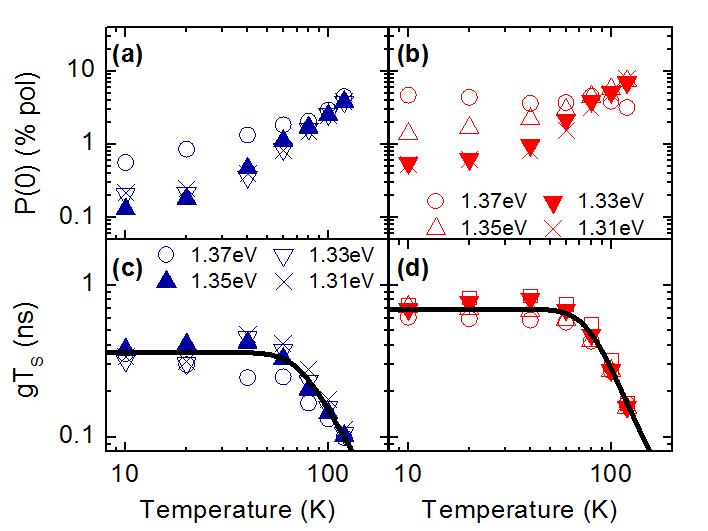}
\caption{
$P(0)$ as a function of temperature with excitation energies of a) 1.59 eV (ExA) and b) 1.45 eV (ExB) at selected photoluminescence (PL) energies.  $gT_s$ as a function of temperature with excitation c) ExA and d) ExB at selected PL energies and fits (solid black lines) as described in the text.  The blue filled up-triangles and red filled down-triangles respectively correspond to the ExA and ExB generated Hanle curves shown in Fig. \ref{f2}.  The size of all data points includes the standard error of the fit values.
}
\label{f3}
\end{figure}

Plots of $P(0)$ and $gT_s$ as functions of temperature and various PL energy for both excitations are shown in Fig. \ref{f3}.  Evaluation of ExA generated PL at 1.35 eV reveals an increase in $P(0)$ from 0.1$\%$ to 4$\%$ polarization and a decrease in $gT_s$ from 0.4 ns to 0.1 ns as temperature increases from 10 K to 120 K.  For ExB generated PL, evaluation at 1.33 eV also reveals an increase of $P(0)$ but from 0.6$\%$ to 7$\%$ and a decrease in $gT_S$ from 0.7 ns to 0.2 ns as temperature increases from 10 K to 120 K.  The temperature dependence of $P(0)$ and $gT_S$, shown in Fig. \ref{f3}, appears to show a threshold for behavior change around 40 K.  The change in $gT_S$ is more apparent than for polarization but both occur around 40 K regardless of excitation.  As PL energy increases, P(0) increases in value below 40 K while $gT_S$ temperature dependence remains unchanged.  This behavior is attributed to hole localization which can be achieved by impurities, Bi clusters, or possibly single Bi atoms.\cite{Francoeur2008}  The $P(0)$ dependence on PL energy might be explained by a correlation of 1.37 eV with free carriers and 1.31 eV to bound carriers.  Recently, a strong nonlinear recombination dependence on Bi composition was reported\cite{Nargelas2011} indicating that As$_{Ga}$ defects play a lesser role in carrier lifetime.  Also observed was thermal activation of hole diffusion which was attributed to strong hole localization.  Furthermore, a report of bismuthide energy gap broadenings\cite{Kudrawiec2012} shows that valence to conduction band transitions are much broader than split-off to conduction band transitions, which is uncommon in III-V alloys and implies unusually strong perturbations.  Bismuth incorporation appears to strongly perturb the valence band and localize holes at isolated Bi atoms or clusters.

The role of the D’yakonov-Perel\cite{D'yakonov1972} and Elliot-Yafet\cite{Elliott1954,Yafet1963}  mechanisms in limiting the measured spin dephasing time should be distinguishable using a temperature dependent power law model \cite{Song2002} and an appropriate assumption for the temperature dependence of the momentum scattering time.  However, the sharp transition in behavior observed at 40 K cannot easily be explained by those mechanisms, and we have insufficient data at higher temperatures to fit the measured spin dephasing time to a temperature dependent power law.  We believe that the dominant temperature dependent effect on $gT_s$ is due to changes in carrier localization.  Below 40 K, holes are localized and therefore undergo minimal scattering; above 40 K, the holes are no longer bound and experience some combination of the above mechanisms.  An Arrhenius function (Eq. \ref{Ar}), analogous to one used in Ref. 20\nocite{Nargelas2011}, was used to fit the temperature dependence of $gT_s$ for both excitation energies, and can be seen as black lines in Fig. \ref{f3}.
\begin{equation}
\label{Ar}
\frac{1}{gT_S}= \alpha e^{-\frac{\Delta E}{kT}}+\frac{1}{g\tau_0}
\end{equation}
$\alpha$ is the pre-exponential factor, T is the sample temperature, k is Boltzmann's constant, $\tau_0$ is the 0 K limit of the dephasing time, and both $g$ and $T_S$ are the same as previously defined.  $\Delta E$ is interpreted as the thermal activation energy for holes with average fit values of 33$\pm$8 meV for ExA and 40$\pm$6 meV for ExB.  These values are similar to the value of 46 meV in Ref. 20\nocite{Nargelas2011}  when a similar model was applied to the temperature dependence of hole diffusion in GaAsBi.  Future work must isolate the contribution of hole localization in order to determine the underlying dephasing mechanisms.

All measurements in this study exhibit an excitation energy dependence which can be explained by changes in carrier concentration.  Spin polarized carriers optically injected in the underlying GaAs could migrate to the bismuthide layer, fill the trap states, and substantially increase the carrier population at the bismuthide band edges.  These spins could partially dephase while in GaAs before crossing the GaAs/GaAsBi interface.  Furthermore, spin polarization of the carriers could scatter as they cross the GaAs/GaAsBi interface, as reported for GaAs/GaNAs \cite{Puttisong2011b} and GaAs/ZnSe \cite{Malajovich2000} heterointerfaces.  The additional spin dephasing of transferred carriers would explain the lower measured polarizations and effective dephasing times for ExA compared to ExB (see Fig. \ref{f3}c,d).  Carrier transfer could also account for the difference in $\Delta E$ values since a higher carrier concentration would lead to a larger number of free carriers, and hence, lower extracted thermal activation energies.

In conclusion, we present measurements of spin dynamics in a dilute bismuthide semiconductor alloy using excitation energies both above and below the GaAs band gap and at temperatures ranging from 10 K to 120 K.  The values of $gT_s$, extracted from Hanle effect measurements, range from 0.8 ns at 40 K to 0.1 ns at 120 K, and are modeled by an Arrhenius function with thermal activation energies of 33$\pm$8 meV for ExA and 40$\pm$6 meV for ExB.  Observations of broad PL spectra below the GaAs bandgap and thermally activated behavior are consistent with existing literature and both phenomena are likely due to hole localization caused by single bismuth or bismuth cluster sites.

This material is based upon work supported by the National Science Foundation under Grant No. ECCS-0844908, and the Materials Research Science and Engineering Center program DMR-1120923.  BP was supported in part by the Graduate Student Research Fellowship under Grant No. DGE 1256260.  VS acknowledges support from the Air Force Office of Scientific Research (Award No. FA9550-12-1-0258) and the Office of Naval Research (Award No. N00014-12-1-0519). GV was supported in part by a Fulbright Foreign Student Fellowship. GV and RSG were supported in part by NSF DMR Grant No. 1006835.

\bibliographystyle{aipnum4-1}
\bibliography{References}

\end{document}